\documentclass{article}

\def\OR{\vee}
\def\AND{\wedge}
\def\goesto{\rightarrow}
\def\implies{\rightarrow}

\newtheorem{definition}{Definition}
\newtheorem{theorem}{Theorem}
\newtheorem{lemma}{Lemma}
\def\beginproof{\noindent{\bf Proof.}\quad}
\def\endproof{$\clubsuit$}

\begin{document}
\title{Geometric properties of satisfying assignments of random $\epsilon$-1-in-$k$ SAT}

\author{Gabriel Istrate, \\
        eAustria Research Institute \\
    Bd. V.~P\^{a}rvan 4, cam 045B, \\
    Timi\c{s}oara, RO-300223, Romania,\\
    email: gabrielistrate@acm.org}

\maketitle

\begin{abstract}
We study the geometric structure of the set of solutions of random
$\epsilon$-1-in-$k$ SAT problem  \cite{1-in-k,zdeborova-1}. For
$l\geq 1$, two  satisfying assignments $A$ and $B$ are {\em
$l$-connected} if there exists a sequence of satisfying assignments
connecting them by changing at most $l$ bits at a time.

We first prove that w.h.p. two assignments of a random
$\epsilon$-1-in-$k$ SAT instance are $O(\log n)$-connected,
conditional on being satisfying assignments. Also, there exists
$\epsilon_{0}\in (0,\frac{1}{k-2})$ such that w.h.p. no two
satisfying assignments at distance at least $\epsilon_{0}\cdot n$
form a "hole" in the set of assignments. We believe that this is
true for all $\epsilon
>0$, and thus satisfying assignments of a random 1-in-$k$ SAT
instance form a single cluster.
\end{abstract}
{\bf Keywords:} $\epsilon$-1-in-$k$ SAT, overlaps, random graphs,
phase transition.

\noindent{\bf AMS Categories:} Primary 68Q25, Secondary 82B27. \\
\noindent{\bf ACM Classification:} F.2.2, G.2.

\section{Introduction}

The geometric structure of solutions of random constraint satisfaction problems has lately become a topic of significant interest \cite{achlioptas-frozen}, \cite{cond-mat/0504070/prl}, \cite{krzakala07}, \cite{istrate-clustering}. The motivation is the study of
{\em phase transitions in combinatorial optimization problems} \cite{weigt-hartmann,comp-statphys}, particularly using methods from physics of Spin Glasses such as the so-called {\em replica method} and {\em cavity approach}. These methods, so far without a complete rigorous foundation, are largely responsible for our substantially increased understanding of structural properties of constraint satisfaction problems.

Of special interest are two special cases when the replica method \cite{virasoro-parisi-mezard} applies, those characterized by so-called ``replica symmetry'' or ``one-step replica symmetry breaking''. These assumptions make predictions on (and have implications for) the typical geometry of the set of solutions of a random instance. Specifically, the two assumptions seem to constrain the set of solutions in the following way:
\begin{enumerate}
 \item For problems displaying replica-symmetry, the set of solutions  forms a single cluster. The typical overlap
is concentrated around a single value, and the distribution of overlaps has continuous support.
\item In the presence of one-step replica-symmetry breaking, the solution space is no longer connected, but breaks into a number of clusters.
 These clusters correspond to the emergence of $\Omega(n)$-size {\em mini-backbones},
 sets of variables taking the same value for all solutions in a cluster. The clusters
 do not possess further geometrical structure (hence the ``one-step'' qualifier in "one step RSB"), and are separated by $\Omega(n)$ variable flips.
The distribution of overlaps develops multiple peaks and has discontinuous support.
\end{enumerate}

In this paper we study the geometric structure of the set of solutions of the {\em random 1-in-$k$ SAT problem  \cite{1-in-k}}, and a generalization of this problem from \cite{zdeborova-1}, {\em random $\epsilon$-1-in-$k$ satisfiability}. This latter problem is parameterized by
a real number $\epsilon\in [0,1/2]$, and essentially coincides with 1-in-$k$ SAT for $\epsilon=1/2$. Results in the cited work suggest that for $\epsilon \in (\epsilon_{c},1/2]$, where
$\epsilon_{c}\sim 0.2726$ is the solution of equation $2x^{3}-2x^{2}+3x-1=0$,
$\epsilon$-1-in-$k$ SAT behaves qualitatively ``like $2$-SAT''. In particular, for both problems the threshold location can be predicted in both cases by a ``percolation of contradictory cycles'' argument, and the replica symmetry ansatz is correct.

For 2-SAT we have previously proved \cite{istrate-clustering} two
results supporting replica symmetry: with high probability
satisfying assignments of a random 2-CNF formula with
clause/variable ratio $c<1$ form a single cluster; also the overlap
distribution has continuous support. From the heuristic similarity
of the two problems, we expect similar results to also hold for
$\epsilon$-1-in-$k$ SAT, $\epsilon \in (\epsilon_{c},1/2]$. Though
the replica symmetric approach seems correct \cite{zdeborova-1}, we
cannot rigorously prove such results. Instead we provide some
evidence for them:
\begin{itemize}
\item We first note (Theorem~\ref{rs}) that the replica symmetric picture holds in the subcritical regime of the formula hypergraph.
 \item We show (Theorem~\ref{first-thm}) that for any two given assignments $A,B$ at sufficiently large Hamming distance, with probability $1-o(1)$
$A,B$ are $O(\log n)$-connected (conditional on being satisfying assignments).
\item We show (Theorem~\ref{second-thm}) that with probability $1-o(1)$ (as $n\goesto \infty$) the set of satisfying assignments of a random instance of 1-in-$k$ SAT with clause/variable ratio $\lambda< \frac{1}{{{k}\choose{2}}}$ does {\em not} have holes of size $>\epsilon_{k}n$, for some $\epsilon_{k}>0$.
\end{itemize}

\section{Preliminaries}

\begin{definition} Let $\epsilon \in [0,1/2]$. An instance of the {\em $\epsilon$-1-in-$k$ SAT  problem} is a propositional formula $\Phi$ in clausal form, with exactly $k$ literals in each clause. A {\em satisfying assignment} for instance $\Phi$ is a mapping of variables in $\Phi$ to $\{0,1\}$ such that in each clause of $\Phi$ {\em exactly} one literal is true.
\end{definition}

We will use two related models to {\em the constant probability model} to generate random instances of $\epsilon$-1-in-$k$ SAT.
\begin{enumerate}
\item The {\em counting} model is parameterized by a real number $r>0$. A random instance of $\epsilon$-1-in-$k$ SAT will have $rn$ clauses, out of
which $rn\cdot \epsilon^{i}(1-\epsilon)^{j}$ have $i$ negative and $j$ positive variables (where $i+j=k$).
\item The {\em constant probability} model is parameterized by a probability $p$. A random instance $\Phi$ is obtained by including independently with probability $p\epsilon^{i}(1-\epsilon)^{j}$ each possible clause
with $i$ negative and $j$ positive variables (where $i+j=k$).
\end{enumerate}
Using standard methods (\cite{bollobas-random-graphs}, Chapter 2;
see also a similar issue in \cite{molloy-stoc2002}) the two models
we described above for $\epsilon$-1-in-$k$ SAT are equivalent:

\begin{lemma} Let $r>0$ and let $p=p(n)$ be such that $p\cdot {{n}\choose {k}} = rn$. Let $\Phi_{1}$ be a random instance of $\epsilon$-1-in-$k$ SAT with $rn$ clauses generated according to the counting model, and let $\Phi_{2}$ be a random instance of $\epsilon$-1-in-$k$ SAT, generated according to the constant probability model with probability $p$. Let $B$ be an arbitrary monotone property and $\mu  \in \{0,1\}$. Then:
\[
 \lim_{n\goesto \infty} Prob[\Phi_{1}\models B]=\mu,
\]
iff
\[
 \lim_{n\goesto \infty} Prob[\Phi_{2}\models B]=\mu.
\]
\end{lemma}

In the sequel we will liberally use one model or the other, depending on our goals.

Results in \cite{1-in-k} and \cite{zdeborova-1} imply the fact that for $\epsilon \in (\epsilon_{c},1/2]$ the threshold of satisfiability for the $\epsilon$-1-in-$k$ satisfiability (under the counting model) is located at critical value\footnote{in \cite{zdeborova-1} the result is only stated and proved for $k=3$, but the method outlined there works for any $k\geq 3$}
\[
 r_{k,\epsilon}=\frac{1}{4\epsilon(1-\epsilon)}\cdot \frac{1}{{{k}\choose {2}}}
\]
The corresponding threshold for $\epsilon$-1-in-$k$ SAT under the constant probability model is
\[
 p_{k,\epsilon}=\frac{(k-2)!}{2\epsilon(1-\epsilon)}\cdot n^{1-k}
\]

\begin{definition}
The {\em overlap} of two assignments $A$ and $B$ for a formula
$\Phi$ on $n$ variables, denoted by $overlap(A,B)$, is the fraction
of variables on which the two assignments agree. Formally
$overlap(A,B)=\frac{|\{i:A(x_{i})=B(x_{i})\}|}{n}. $
\end{definition}

The distribution of overlaps is, indeed, the original order
parameter that was originally used to study the phase transition in
random $k$-SAT \cite{monasson:zecchina}.

\begin{definition}
 Let $l\geq 1$ be an integer and let $A,B$ be two satisfying assignments of an instance $\Phi$ of $\epsilon$-1-in-$k$ SAT. Pair $(A,B)$ is called {\em $l$-connected} if there exists a sequence of satisfying assignments $A_{0}, A_{1}, \ldots A_{r}$, $A_{0}=A$, $A_{r}=B$, with $A_{i}$ and $A_{i+1}$ at Hamming distance at most $l$.

\end{definition}

\begin{definition}
 Let $A,B$ be arbitrary assignments for the variables of an instance $\Phi$ of 1-in-$k$ SAT. Pair $(A,B)$ is called
a {\em hole} if:
\begin{enumerate}
 \item $A,B$ are satisfying assignments for $\Phi$.
\item There exists no satisfying assignment $C$ with $d_{H}(A,C)+d_{H}(C,B)=d_{H}(A,B)$ (where $d_{H}$ is the Hamming distance).
\end{enumerate}
The number $\lambda=d_{H}(A,B)$ is called {\em the size of hole $(A,B)$}.
\end{definition}

\section{Results}

First, we prove that for low enough clause/variable ratios the set of satisfying assignments of a random instance of 1-in-$k$ SAT behaves in the way predicted by the replica symmetry ansatz:

\begin{theorem}\label{rs}
 Let $k\geq 3$ and $c<1/k(k-1)$. Then there exists $\gamma>0$ such that, with probability $1-o(1)$ (as $n\goesto \infty$), a random instance of 1-in-$k$ SAT with $n$ variables and $cn$ clauses has all its satisfying assignments
$\gamma\log(n)$-connected.
\end{theorem}

We believe (and would like to prove) that the result in Theorem~\ref{rs} is valid for values of $c$ up to $2/k(k-1)$ (the satisfiability threshold of 1-in-$k$ SAT \cite{1-in-k}). We cannot prove this statement. Instead, we prove a result that implies a weaker claim for 1-in-$k$ SAT but is valid, more generally, for $\epsilon$-1-in-$k$ SAT:

\begin{theorem}\label{first-thm}
 Let $0\leq \epsilon \leq \frac{1}{2}$, let $c<1$, let $\Phi$ be a random instance of $\epsilon$-1-in-$k$ SAT with clause/variable ratio $\frac{1}{\max[4\epsilon(1-\epsilon),\epsilon^2+(1-\epsilon)^2]}\cdot \frac{c}{{{k}\choose {2}}}$,
and let $(A_{n},B_{n})\in \{0,1\}^{n}\times \{0,1\}^{n}$ such that
\begin{equation}\label{cond}
2\cdot [overlap(A_{n},B_{n})(1-\epsilon)]^{k-2}\leq 1.
\end{equation}

Then there exists $\lambda_{c,\epsilon}>0$ such that
\[
 \Pr[(A_{n},B_{n})\mbox{ are not }\lambda_{c,\epsilon}\cdot\log(n)\mbox{-connected }|\mbox{ }A_{n},B_{n}\models \Phi\}<1/n.
\]
for large enough $n$.
\end{theorem}

In other words, every {\em single pair} of assignments is likely to be $O(\log n)$-connected, conditional on being a pair of satisfying assignments, {\em and being far enough}. The remarkable thing about condition~(\ref{cond}) is that it depends on $\epsilon$ and $k$ but {\bf not} $c$. For certain values of $\epsilon$
(we specifically believe this is the case in the region $[0,\epsilon_{c})$)
it might simply signal the fact that there are no satisfying assignments of a certain overlap. This is not a problem for $\epsilon = 1/2$ (i.e. for the
1-in-$k$ SAT), since the condition~(\ref{cond}) is trivially satisfied for every overlap value. For this problem, the results in
Theorems~\ref{rs} and \ref{first-thm} are highly reminiscent of the results for 2-SAT in \cite{istrate-clustering}. On the other hand for any $c<1$ and all $q\in (0,1)$ a random 2-CNF formula has w.h.p. two satisfying assignments of overlap approximately $q$. Despite 2-SAT and 1-in-$k$ SAT being similar in other ways (see e.g. \cite{aimath04}), the corresponding statement is {\em not} true for 1-in-$k$ SAT:

\begin{theorem}\label{diff}
For any $c>0$ there exists $q_{c}\in (0,1)$ such that w.h.p. a random instance of 1-in-$k$ SAT with clause/variable ratio $c$ has, with probability $1-o(1)$ no satisfying assignments of overlap less than $q_{c}$.
\end{theorem}

We next consider an alternative approach to characterizing the
geometry of satisfying assignments of 1-in-$k$ SAT by studying the
existence of {\em holes} inbetween such assignments. For other
problems, e.g. $k$-SAT, $k\geq 9$, that display clustering the set
of satisfying assignments has large holes. Indeed
\cite{cond-mat/0504070/prl}, for certain values of
$q_{1}<q_{2}<q_{3}$ and $c>0$, a random instance of $k$-SAT of
constraint density $c$ will have, with high probability, satisfying
assignments of overlap $q_{3}$, but no satisfying assignments of
overlap $\lambda$, $q_{1}\leq \lambda \leq q_{2}$. Consider $A,B$
two satisfying assignments of overlap $q_{3}$. Then the set of
assignments $C$ between $A,B$ contains a hole of size at least
$(q_{2}-q_{1})n$.

We would like to stat that for any $\lambda >0$ a random instance $\Phi$ of 1-in-$k$ SAT as in Theorem~\ref{first-thm} has {\em no} hole of size at least $\lambda \cdot n$. We cannot, however prove this result
(we leave it as an intriguing open problem). Instead we prove a weaker result:

\begin{theorem}\label{second-thm}
 For any $k\geq 3$ there exists $\epsilon_{k}\in (0,1/k-1)$ such that with probability $1-o(1)$ (as $n\goesto \infty$) a random instance of 1-in-$k$ SAT of clause/variable ratio $c<1/{{k}\choose {2}}$ has no holes of size $\geq \epsilon_{k}\cdot n$.
\end{theorem}

\section{Proofs}

\subsection{Proof of Theorem~\ref{rs}}

First, note that location $c=1/k(k-1)$ in Theorem~\ref{rs} is the phase transition location for the random $k$-uniform hypergraph \cite{phase-transition-random-hypergraph}. For smaller values of $c$, by results in \cite{phase-transition-random-hypergraph} there exists $\gamma>0$ such that w.h.p. the largest connected component of $H$ has size no larger than $\gamma\log(n)$.

This argument immediately implies the desired result. Indeed, let $P,Q$ be two arbitrary satisfying assignments, and let $(P_{1},Q_{1})$, $(P_{2},Q_{2})$, $\ldots, (P_{v},Q_{v})$ represent the restrictions of $P$ and $Q$ on the connected components of $\Phi$ on which $P\neq Q$. One can obtain a path from $P$ to $Q$ by starting at $P$ and then obtain the next satisfying assignments by replacing $P_{i}$ by $Q_{i}$ for $i=1,\ldots, v$. In this way we are constructing satisfying assignments for $\Phi$, since we change assignments consistently on connected components of the formula hypergraph. We are changing at most $\gamma\log(n)$ values at a time, since this is the upper bound on the component size of $H$.

\subsection{Proof of Theorem~\ref{diff}}

We prove the theorem by a simple first moment bound. We will work with the
constant probability model.
\begin{definition}
 Let $\Phi$ be a formula. A {\em cover of $\Phi$} is a set of variables $W$ such that every clause of $\Phi$ contains at least one variable in $W$.
\end{definition}

The theorem now follows from the following two lemmas:

\begin{lemma}
 Let $A$,$B$ be satisfying assignments of an instance $\Phi$ of 1-in-$k$ SAT. Then the set $\{x:A(x)=B(x)\}$ is a cover of $\Phi$.
\end{lemma}
\beginproof

Suppose this was not the case, and there exists a clause $C$ of $\Phi$ consisting entirely of variables in the set $\{x:A(x)\neq B(x)\}$. Then clause $C$ has two satisfying assignments at distance $k$. But this is not possible, since all satisfying assignments of a given 1-in-$k$ clause have Hamming distance two.
\endproof

\begin{lemma}
 For any $c>0$ there exists a $q_{c}>0$ such that a random instance of 1-in-$k$ SAT of clause/variable ratio $c$ has, w.h.p. no cover of size at most $q_{c}n$.
\end{lemma}
\beginproof
Let $\lambda <1/2$.
The probability that $\Phi$ has a cover of size $i\leq \lambda n$ is at most
\begin{eqnarray*}
 & & \sum_{i=1}^{\lambda n} {{n}\choose {i}} (1-p)^{{{n-i}\choose {k}}} \leq
 \lambda n \cdot (1-p)^{{{n(1-\lambda)}\choose {k}}}\cdot \Big[\sum_{i=1}^{\lambda n} {{n}\choose {i}}\Big]\leq \\ & \leq &
(\lambda n)^2\cdot e^{-p{{n(1-\lambda)}\choose {k}}}\cdot {{n}\choose {\lambda n}}\leq (1+w)\cdot (\lambda n)^2\cdot e^{-p{{n(1-\lambda)}\choose {k}}}\cdot \Big(\frac{1}{\lambda^{\lambda}(1-\lambda)^{1-\lambda}}\Big)^{n}\cdot \\
& \cdot & \frac{1}{\sqrt{2\pi\lambda(1-\lambda)n}}.
\end{eqnarray*}
for some $w>0$ (we have applied the fact that $\lambda < 1/2$ and Stirling's formula)
So the probability is at most
\begin{eqnarray*}
 & & \frac{(1+w)(\lambda n)^2}{\sqrt{2\pi\lambda(1-\lambda)n}}\cdot e^{-p(n(1-\lambda))^{k}/k!+ n [\lambda\ln(1/\lambda)+(1-\lambda)\ln(1/(1-\lambda))]}= \\ & = &  \frac{(1+w)(\lambda n)^2}{\sqrt{2\pi\lambda(1-\lambda)n}}\cdot e^{-n[c(1-\lambda))^{k}- \lambda\ln(1/\lambda)-(1-\lambda)\ln(1/(1-\lambda))]}
\end{eqnarray*}

Since $c>0$ and $
 \lim_{\lambda\goesto 0}\lambda\ln(1/\lambda)-(1-\lambda)\ln(1/(1-\lambda))=0
$,
there exists $q_{c}>0$ such that for $\lambda<q_{c}$, $c(1-\lambda))^{k}- \lambda\ln(1/\lambda)+(1-\lambda)\ln(1/(1-\lambda))>0$. Thus, for $q<q_{c}$
the probability that a random instance of 1-in-$k$ SAT has a cover of size at most $qn$ is exponentially small.
\endproof

\subsection{Proof of Theorem~\ref{first-thm}}
For a pair of assignments $(A,B)$ define
\begin{eqnarray*}
V_{0}=\{x:A(x)=B(x)=0\},
V_{1}=\{x:A(x)=0,B(x)=1\},\\
V_{2}=\{x:A(x)=1,B(x)=0\},
V_{3}=\{x:A(x)=B(x)=1\}.
\end{eqnarray*}

Pair $(A,B)$ has {\em type $(a,b,c,d)$} if
$|V_{0}|=a,|V_{1}|=b,|V_{2}|=c,|V_{3}|=d$. Also denote
$\alpha = a/n, \beta=b/n, \gamma=c/n, \delta =d/n$.

Conditioning  on $A,B$ being satisfying assignments, define a graph $H$ on the set of variables in $A\neq B$ as follows: $x$ and $y$ are connected if there exists a clause $C$ of $\Phi$ consisting of $k-2$ literals whose variables are from $V_{0}\cup V_{3}$ and $x,y$. Since both $A$ and $B$ must be satisfying assignments, only four combinations are possible for the literal combination present in $C$:
\begin{enumerate}
\item ($x,y\in C$ or $\overline{x},\overline{y}\in C$) and $A(x)\neq A(y)$, or
\item ($x,\overline{y}\in C$ or $\overline{x},y\in C$) and $A(x)=A(y)$.
\end{enumerate}

We can rewrite conditions (1) and (2) as
\begin{enumerate}
\item ($x,y\in C$ or $\overline{x},\overline{y}\in C$) and $(x\in V_{1}\AND y\in V_{2})\OR (x\in V_{2}\AND y\in V_{1})$, or
\item ($x,\overline{y}\in C$ or $\overline{x},y\in C$) and $(x,y\in V_{1})\OR (x,y\in V_{2})$.
\end{enumerate}

To summarize this discussion, there are four types of clauses that
imply the existence of an edge $(x,y)$ in graph $H$. They are
described in the table from Figure 1. The semantics of columns in
the table is the following: first column (type) lists the four types
of clauses, labeled $C_{1}$ to $C_{4}$. Columns labeled $V_{0}$ to
$V_{3}$ contain two numbers. The first one is the number of literals
of the given clause type that are in the set $V_{j}$. The second
number (in square brackets) lists the number of negated variables in
the set $V_{j}$. Column labeled ``number'' computes the total number
of clauses of type $C_{i}$. The column labeled ``Probability'' lists
the probability that a fixed clause of type $C_{i}$ be in $\Phi$.

\begin{figure}\label{table}
\begin{center}
 \begin{tabular}{|c|c|c|c|c|c|c|}
\hline\hline
type & $V_{0}$ $(a)$ & $V_{3}$ $(d)$ & $V_{1}$ $(b)$ & $V_{2}$ $(c)$ & number & probability \\
\hline
$C_{1}$ & $k-i-2$ $[0]$ & $i$ $[i]$ & $1$ $[0]$ & $1$ $[0]$ & ${{a}\choose {k-i-2}}{{d}\choose {i}}$ & $p\epsilon^{i}(1-\epsilon)^{k-i}$ \\
\hline
$C_{2}$ & $k-i-2$ $[0]$ & $i$ $[i]$ & $1$ $[1]$ & $1$ $[1]$ & ${{a}\choose {k-i-2}}{{d}\choose {i}}$ & $p\epsilon^{i+2}(1-\epsilon)^{k-i-2}$ \\
\hline
$C_{3}$ & $k-i-2$ $[0]$ & $i$ $[i]$ & $2$ $[1]$ & $0$ $[0]$ & $2{{a}\choose {k-i-2}}{{d}\choose {i}}$ & $p\epsilon^{i+1}(1-\epsilon)^{k-i-1}$ \\
\hline
$C_{4}$ & $k-i-2$ $[0]$ & $i$ $[i]$ & $0$ $[0]$ & $2$ $[1]$ & $2{{a}\choose {k-i-2}}{{d}\choose {i}}$ & $p\epsilon^{i+1}(1-\epsilon)^{k-i-1}$ \\
\hline\hline
 \end{tabular}
\end{center}
\caption{The four types of clauses leading to an edge $(x,y)$ in graph $H$}
\end{figure}

The probability that an edge is present in graph $H$ is the same for all pairs $(x,y)$ such that $A(x)=A(y)$. Similarly the probability that an edge is present in graph $H$ is the same for all pairs $(x,y)$ such that $A(x)\neq A(y)$. We denote by $\mu_{=}=\mu_{=}(n,a,b,c,d)$ and $\mu_{\neq}=\mu_{\neq}(n,a,b,c,d)$ these two probabilities.

\begin{eqnarray*}
 & \mu_{=} & \leq p\cdot \sum_{i=0}^{k-2}{{a}\choose {k-i-2}}{{d}\choose {i}}\cdot \Big[\epsilon^{i}(1-\epsilon)^{k-i}+ \epsilon^{i+2}(1-\epsilon)^{k-i-2}\Big]
\end{eqnarray*}

\begin{eqnarray*}
 & \mu_{\neq} & \leq p\cdot \Big\{\sum_{i=0}^{k-2}{{a}\choose {k-i-2}}{{d}\choose {i}}\cdot \Big[2\epsilon^{i+1}(1-\epsilon)^{k-i-1}+ 2\epsilon^{i+1}(1-\epsilon)^{k-i-1}\Big] \Big\}
\end{eqnarray*}

Applying inequality ${{a}\choose {i}}\leq \frac{a^{i}}{i!}$ and rewriting the second term of the previous inequalities we get
\begin{eqnarray*}
 &\mu_{=} & \leq \frac{p[\epsilon^2+(1-\epsilon)^2]}{(k-2)!} \cdot \Big\{\sum_{i=0}^{k-2}{{k-2}\choose {i}}\cdot a^{k-i-2}d^{i}\cdot \epsilon^{i}(1-\epsilon)^{k-i-2}\Big\}= \\ & = & \frac{p[\epsilon^2+(1-\epsilon)^2]}{(k-2)!} \cdot [a(1-\epsilon)+d\epsilon]^{k-2}
\end{eqnarray*}
\begin{eqnarray*}
& \mu_{\neq}& \leq  \frac{4p\epsilon(1-\epsilon)}{(k-2)!} \cdot \Big\{\sum_{i=0}^{k-2}{{k-2}\choose {i}}\cdot a^{k-i-2}d^{i}\cdot \epsilon^{i}(1-\epsilon)^{k-i-2}\Big\}= \\
& = & \frac{4p\epsilon(1-\epsilon)}{(k-2)!} \cdot [a(1-\epsilon)+d\epsilon]^{k-2}
\end{eqnarray*}

The equation $\epsilon^2+(1-\epsilon)^2=4\epsilon(1-\epsilon)$ has a solution $\epsilon_{0}=\frac{3-\sqrt{3}}{6}\sim 0.2113...$. For $\epsilon\in (\epsilon_{0},1/2]$ we have $\epsilon^2+(1-\epsilon)^2<4\epsilon(1-\epsilon)$.

For $p=\lambda\cdot k!\cdot n^{1-k}$, with $\lambda=\frac{c}{4\epsilon(1-\epsilon)}\cdot\frac{2}{k(k-1)}$, with  $c<1$
we have  $\max(\mu_{=},\mu_{\neq})=\frac{2c}{n}[\alpha(1-\epsilon)+\delta\epsilon]^{k-2}\leq \frac{c}{n}\cdot 2\cdot [overlap(A,B)(1-\epsilon)]^{k-2}\leq \frac{c}{n}$. It follows that the
graph $H$ has all its connected components of size at most $\lambda_{c}\log n$, where \cite{janson-random-graphs}
 \begin{equation}
\lambda_{c}= \frac{3}{(1-c)^2}
\end{equation}
By the discussion of clause types in Figure 1, edges of type (1)
correspond to a constraint $x\neq y$ between a variable in $V_{1}$
and one in $V_{2}$, while clauses of type (2) correspond to a
constraint $x=y$ between two variables, both in $V_{1}$ or both in
$V_{2}$. $H$ does not contain contradictory cycles, since we have
conditioned on $A,B$ being satisfying assignments.

It is easy to see that setting one value of a given variable in $H$ uniquely determines the values on its whole connected component. Similarly, different values of $x$ lead to opposite assignments on the connected component of $x$. Given the small size of the connected components, the statement of the theorem immediately follows.
\endproof
\subsection{Proof of Theorem~\ref{second-thm}}
We first prove a simple result about the connectivity of a random graph that we will use in the sequel.

\begin{lemma}\label{connected}
 Let $0<c<1$ and let $G$ be a random graph from $G(n,c/n)$. Then
\[
 Pr[G\mbox{ is connected }]\leq \frac{c^{n-1}}{n}.
\]
\end{lemma}
\beginproof
 There are $u=n^{n-2}$ labeled trees on the set of vertices of $G$. Denote by $T_{1}, \ldots T_{u}$ the edge sets of these trees, and by $W_{i}$ the event ``$T_{i}\subseteq E[G]$''. It is easy to see that $G$ is {\em not} connected if and only if $\bigwedge_{i=1}^{u} \overline{W_{i}}$.

By Janson's inequality \cite{probabilistic-method} \[\Pr[\bigwedge_{i=1}^{u} \overline{W_{i}}]\geq \prod_{i=1}^{u} \Pr[\overline{W_{i}}] =
(1 - (\frac{c}{n})^{n-1})^{u}.
\]

So, by applying inequality $1-(1-x)^{a}\leq ax$ we get:
\[
 \Pr[G\mbox { connected }]\leq 1-(1 - (\frac{c}{n})^{n-1})^{n^{n-2}}\leq \frac{c^{n-1}}{n},
\]
\endproof

We will work with the constant probability model. Each clause
will be included in formula $\Phi$ with probability $p$,  where $p\cdot 2^{k}\cdot {{n}\choose {k}}= \lambda \cdot \frac{1}{{{k}\choose {2}}}n$, with $\lambda <1$.

\begin{lemma}
 The probability that there exist two satisfying assignments $A$ and $B$ of overlap $i$ that form a hole is at most
\[\frac{{{n}\choose {i}}\cdot 2^{i}\cdot [2^{2-k}\cdot
(\frac{i}{n})^{k-2}\cdot(1-\frac{i}{n})]^{n-i-1}}
{(n-i)}\cdot e^{-\frac{\lambda n}{{{k}\choose {2}}}
[1-\frac{k {{i}\choose {k}}+2{{i}\choose {k-2}}
{{n-i}\choose {2}}}{{2^k}\cdot{{n}\choose {k}}}]}
 \]

\end{lemma}
\beginproof

For two assignments $A,B$ of overlap $i$ to be satisfying assignments of a formula $\Phi$, all clauses of $\Phi$ must fall into one of the following two categories:
\begin{enumerate}
 \item Clause $C$ contains $k-1$ literals from $A=B=0$ and one literal from $A=B=1$.
\item Clause $C$ contains $k-2$ literals from $A=B=0$ and two literals from $A\neq B$, of opposite signs in $A$.
\end{enumerate}
There are $k\cdot {{i}\choose {k}}$ clauses of the first
type and $2{{i}\choose {k-2}}\cdot {{n-i}\choose {2}}$
clauses of the second type. So the probability that all
clauses of $\Phi$ fall into these two categories is
\begin{eqnarray*}& & (1-p)^{2^{k}\cdot {{n}\choose {k}}-k\cdot {{i}\choose
{k}}-2{{i}\choose {k-2}}{{n-i}\choose {2}}}\leq e^{-p\cdot [2^{k}{{n}\choose {k}}-k\cdot {{i}\choose
{k}}-2{{i}\choose {k-2}}{{n-i}\choose {2}}]} = \\ & = & e^{-\frac{\lambda n}{{{k}\choose {2}}}
[1-\frac{k {{i}\choose {k}}+2{{i}\choose {k-2}}
{{n-i}\choose {2}}}{{2^k}\cdot{{n}\choose {k}}}]}  .\end{eqnarray*}

The probability is at most ${{n}\choose {i}}\cdot 2^{i}$ times the probability that giving specific values to $i$ variables we simplify the formula $\Phi$ to one for which the following graph $H$ is connected: two variables $x,y\in \{\lambda:A(\lambda)\neq B(\lambda)\}$ are connected if there exists a clause $C$ of $\Phi$ that contains both of them (and no other variables in that set).

This probability is at most $2\cdot {{i}\choose {k-2}}\cdot p$. So an uper bound for the probability is
\begin{eqnarray*}
 & 2 &\cdot \frac{i^{k-2}}{(k-2)!}\cdot 2^{-k}\cdot \frac{k!}{n^{k}}\cdot \lambda \cdot \frac{1}{{{k}\choose {2}}}n = \frac{(\frac{i}{2n})^{k-2}\cdot\lambda}{2n} = \\ & = &
\frac{2\cdot (\frac{i}{2n})^{k-2}\cdot(1-\frac{i}{n})\cdot \lambda}{n-i} \leq \frac{2^{2-k}\cdot   (\frac{i}{n})^{k-2}\cdot(1-\frac{i}{n})}{n-i}.
\end{eqnarray*}

Since connectivity is an increasing property, applying Lemma~\ref{connected}, the probability that $H$ is connected is at most $\frac{{[2^{2-k}\cdot\lambda\cdot   (\frac{i}{n})^{k-2}\cdot(1-\frac{i}{n})]}^{n-i-1}}{ (n-i)}$.
\endproof

Let $\alpha = i/n$. Then the upper bound in the result above reads:
\begin{eqnarray*}
 \frac{{{n}\choose {\alpha\cdot n}}\cdot 2^{\alpha\cdot n}\cdot [2^{2-k}\cdot \lambda\cdot  (\frac{\alpha\cdot n}{n})^{k-2}\cdot(1-\frac{\alpha\cdot n}{n})]^{n-\alpha\cdot n-1}} {n(1-\alpha)}\cdot
e^{-\frac{\lambda n}{{{k}\choose {2}}}\cdot
[1-\frac{k\alpha^k+k(k-1)\alpha^{k-2}(1-\alpha)^2}{2^k}]}
\end{eqnarray*}

Applying Stirling's formula for the factorial, the above expression simplifies to
\begin{eqnarray*}
 & & \frac{(\frac{n}{e})^{n}\sqrt{2\pi n}\cdot 2^{\alpha\cdot n}\cdot [2^{2-k}\lambda(\frac{\alpha
n}{n})^{k-2}(1-\frac{\alpha
n}{n})]^{n-\alpha n-1}\cdot
e^{-\frac{\lambda n}{{{k}\choose {2}}}\cdot
[1-\frac{k\alpha^k+k(k-1)\alpha^{k-2}(1-\alpha)^2}{2^k}]} }{(\frac{\alpha \cdot
n}{e})^{\alpha \cdot n}\sqrt{2\pi\alpha n}\cdot
(\frac{(1-\alpha) \cdot n}{e})^{(1-\alpha) \cdot
n}\sqrt{2\pi(1-\alpha)n} \cdot n(1-\alpha)} = \\
& = & \theta(1)\cdot \frac{2^{\alpha\cdot n}\cdot [2^{2-k}\lambda \alpha ^{k-2}(1-\alpha)]^{n(1-\alpha)}\cdot
e^{-\frac{\lambda n}{{{k}\choose {2}}}\cdot
[1-\frac{k\alpha^k+k(k-1)\alpha^{k-2}(1-\alpha)^2}{2^k}]} }{\alpha^{\alpha n}\sqrt{\alpha}\cdot (1-\alpha)^{(1-\alpha)n}\sqrt{2\pi(1-\alpha)n} \cdot n(1-\alpha)\cdot (\alpha/2)^{k-2}\lambda(1-\alpha)}= \\
& = & \theta(1)\cdot  \frac{n^{-3/2}}{\alpha^{k-3/2}\lambda(1-\alpha)^{5/2}}\cdot \\
& \cdot & \Big\{\frac{2^{\alpha}[(\alpha/2) ^{k-2}\lambda(1-\alpha)]^{(1-\alpha)}
e^{-\lambda/{{k}\choose {2}}} (1-\frac{k\alpha^k+k(k-1)\alpha^{k-2}(1-\alpha)^2}{2^k})
}
{\alpha^{\alpha}\cdot (1-\alpha)^{1-\alpha}}\Big\}^{n} = \\
& = & \theta(1)\cdot \frac{n^{-3/2}}{\alpha^{k-3/2}\lambda(1-\alpha)^{5/2}} \cdot f_{k}(\alpha)^{n},
\end{eqnarray*}
where
\[
 f_{k}(x)= \lambda^{1-x}\cdot (x/2)^{(k-2)(1-x)-x}\cdot e^{-\lambda  (1-\frac{k x^k+k(k-1)x^{k-2}(1-x)^2}{2^k})/{{k}\choose {2}}}
\]
and the $\theta(1)$ factor does {\em not} depend on $\alpha$ or $\lambda$.

Let
\begin{eqnarray*}
 & g_{k}(x) & =  \ln (f_{k}(x))= \\
& = & (1-x)\ln \lambda + [(k-2)(1-x)-x]\cdot \ln (x/2) - \\
& - & \frac{\lambda}{{{k}\choose {2}}}(1-\frac{k x^k+k(k-1)x^{k-2}(1-x)^2}{2^k}).
\end{eqnarray*}

For $x\in (0,\frac{k-1}{k-2}]$, since $\lambda<1$, $\ln \lambda<0$
and $1-x>0$. Also $\ln(x/2)<0$ while $(k-2)(1-x)-x>0$. Finally,
$k(k-1)x^{k-2}(1-x)^{2}\leq k(k-1)/2$ (since $x<1$ and $x(1-x)\leq
1/4$. Since $k+k(k-1)/2=\frac{k(k+1)}{2}<2^{k}$ (since $k\geq 3$),
we infer that the last term is positive.

The conclusion of this argument is that
$x\in (0,\frac{k-1}{k-2}]\implies g_{k}(x)<0$.

On the other hand
$g_{k}(1)=\ln 2 - \frac{\lambda}{{{k}\choose {2}}} \Big(1-\frac{k}{2^{k}}\Big)>\ln 2 - \frac{1}{{{k}\choose {2}}}>0$, since $k\geq 3$. Thus the equation
 $g_{k}(x)=0$ has a (smallest) root $x_{k}\in (\frac{k-1}{k-2},1)$.

For $\alpha < x_{k}$, $f(\alpha)<1$ and the upper bound is asymptotically equal to zero.
\endproof
\section{Conclusions}

Theorem~\ref{rs} connects the percolation of the giant component in the random formula hypergraph to the existence of a single cluster of satisfying assignments. Of course, since the phase transition in 1-in-$k$ SAT is determined \cite{1-in-k} by a ``giant component`` phenomenon in a directed
version of the formula hypergraph, the main open question raised by this work is to prove that the statement of Theorem~\ref{rs} holds up to critical
threshold $c=\frac{2}{k(k-1)}$. Theorem~\ref{first-thm} provides further evidence that this might be true.

We believe that it might be possible to prove this statement using a more robust generalization of the notion of ''hole`` in the set of satisfying assignments.

\section*{Acknowledgments}

I thank Romeo Negrea for useful discussions. This work has been
supported by a Marie Curie International Reintegration Grant within
the 6th European Community Framework Programme and by a
PN-II/"Parteneriate" grant from the Romanian CNCSIS.

\bibliographystyle{gCOM}

\begin{thebibliography}{10}
\providecommand{\url}[1]{\texttt{#1}}
\providecommand{\urlprefix}{URL }

\bibitem{achlioptas-frozen}
D. Achlioptas and F. Ricci-Tersenghi, \emph{On the solution space
geometry of
  random constraint satisfaction problems}, in \emph{Proceedings of the 36th
  ACM Symposium on Theory of Computing}, 2006, pp. 130--139.

\bibitem{1-in-k}
D. Achlioptas, A. Chtcherba, G. Istrate,  and C. Moore, \emph{The
phase
  transition in 1-in-$K$ {SAT} and {NAE3SAT}}, in \emph{Proceedings of the 12th
  ACM-SIAM Symposium on discrete algorithms}, 2001.

\bibitem{probabilistic-method}
N. Alon, P. Erd\H{o}s,  and J. Spencer, \emph{The probabilistic
method}, 2nd
  ed., John Wiley and Sons (1992).

\bibitem{bollobas-random-graphs}
B. Bollob\'{a}s, \emph{Random graphs}, 2nd ed., Cambridge University
Press, 2001.

\bibitem{weigt-hartmann}
A. Hartmann and M. Weigt, \emph{Phase transitions in combinatorial
optimization
  problems}, Wiley-VCH (2005).

\bibitem{istrate-clustering}
G. Istrate, \emph{Satisfiability of boolean random constraint
satisfaction:
  Clusters and overlaps}, Journal of Universal Computer Science 13 (2007), pp.
  1655--1670.

\bibitem{aimath04}
G. Istrate, A. Percus,  and S. Boettcher, \emph{Spines of random
constraint
  satisfaction problems: Definition and connection with computational
  complexity}, Annals of Mathematics and Artificial Intelligence 44 (2005), pp.
  353--372.

\bibitem{janson-random-graphs}
S. Janson, T. Luczak,  and A. Ruczinski, \emph{Random Graphs}, John
Wiley \&
  Sons (2000).

\bibitem{krzakala07}
F. Krzakala, A. Montanari, F. Ricci-Tersenghi, G. Semerjian,  and L.
Zdeborova,
  \emph{Gibbs states and the set of solutions of random constraint satisfaction
  problems}, Proceedings of the National Academy of Sciences 104 (2007), pp.
  10318--10323.

\bibitem{cond-mat/0504070/prl}
M. M\'{e}zard, T. Mora,  and R. Zecchina, \emph{Clustering of
solutions in the
  random satisfiability problem}, Physical Review Letters 94 (2005).

\bibitem{virasoro-parisi-mezard}
M. M\'{e}zard, G. Parisi,  and M. Virasoro, \emph{Spin glass theory
and
  beyond}, World Scientific (1987).

\bibitem{molloy-stoc2002}
M. Molloy, \emph{Models for random constraint satisfaction
problems}, in
  \emph{Proceedings of the 32nd ACM Symposium on Theory of Computing}, 2002.

\bibitem{monasson:zecchina}
R. Monasson and R. Zecchina, \emph{Statistical mechanics of the
random
  $k$-{SAT} model}, Physical Review E 56 (1997), p. 1357.

\bibitem{comp-statphys}
A. Percus, G. Istrate,  and C. Moore (eds.), \emph{Computational
Complexity and
  Statistical Physics}, Oxford University Press (2006).

\bibitem{zdeborova-1}
J. Raymond, A. Sportiello,  and L. Zdeborov\'{a}, \emph{The phase
diagram of
  random 1-in-3 satisfiability}, Phys. Rev. E 76 (2007).

\bibitem{phase-transition-random-hypergraph}
J. Schmidt-{P}ruznan and D. Shamir, \emph{Component structure in the
evolution
  of random hypergraphs}, Combinatorica 5 (1985), pp. 81--94.

\end{thebibliography}

\end{document}